\documentclass[aps,prd,twocolumn]{revtex4}
\usepackage{amsmath}
\usepackage{graphicx}

\def\mpl{M_{\rm Pl}}
\def\ex{\epsilon}
\def\Lx{\Lambda}

\def\elt{\tilde{\ell}}
\def\tt{\tilde{t}}
\def\lef{\lambda_{\rm eff}}
\def\calm{{\cal M}}

\newcommand{\rht}{\tilde{\rho}}

\begin{document}

\title{Late acceleration and $w=-1$ crossing 
in induced gravity}
\author{Pantelis S. Apostolopoulos$^1$\footnote{Email address: vdfspap8@uib.es; papost@phys.uoa.gr} and Nikolaos Tetradis$^2$\footnote{Email address: ntetrad@phys.uoa.gr}}
\date{\today}
\address{$^1$Departament de F\'isica, Universitat de les Illes Balears, Cra. Valldemossa Km 7.5, E-07122 Palma de Mallorca, Spain\\\\\\
$^2$University of Athens, Department of Physics, University Campus, Zographou 157 84, Athens, Greece}
\begin{abstract}
We study the cosmological evolution on a brane with induced gravity within
a bulk with arbitrary matter content. We consider
a Friedmann-Robertson-Walker brane, invariantly characterized
by a six-dimensional group of isometries. We derive the effective
Friedmann and Raychaudhuri equations. We show that the Hubble expansion
rate on the brane depends on 
the covariantly defined integrated mass in the 
bulk, which determines the energy density of the generalized 
dark radiation.
The Friedmann equation has two branches, distinguished by the
two possible values of the parameter $\ex=\pm 1$. The branch with 
$\ex=1$ is characterized by an effective cosmological constant and 
accelerated expansion for low energy densities.
Another remarkable feature is 
that the contribution from the generalized 
dark radiation appears with a negative
sign. As a result, the presence of the bulk 
corresponds to an effective negative energy density on the brane, without
violation of the weak energy condition.
The transition from a period of domination of the matter energy density by 
non-relativistic brane matter to domination by the generalized dark radiation 
corresponds to a crossing of the phantom divide $w=-1$.
\end{abstract}

\maketitle


\setcounter{equation}{0}

\section{Introduction}

A brane theory with an induced gravity term in the action, characterized by a
length scale $r_c$,
has several novel features. In the simplest example, the DGP model \cite{DGP}, 
the brane tension $V$ and bulk cosmological constant $\Lx$ are neglected. 
Specific examples with 
an induced gravity term can be obtained in string theory, and are 
common in holographic descriptions \cite{inducedrs,kohlp,holo,kiritsish}. 
Despite possible problems at large distances \cite{strongcoupl}, 
the DGP model predicts an interesting cosmological evolution \cite{deffayet}.
The Friedmann equation has two branches, distinguished by the
two possible values of the parameter $\ex=\pm 1$. The branch with 
$\ex=1$ is characterized by an effective cosmological constant and 
accelerated expansion for low energy density of the brane matter
\cite{acceldgp}.
The generalization of the model for non-zero $V$ and $\Lx$ 
displays the same two branches \cite{inducedrs}. For $\ex=-1$, the 
limit $r_c \to 0$ reproduces the cosmological evolution 
of the Randall-Sundrum model \cite{rs,binetruy}, in which the
effective cosmological constant is zero. 
The branch with $\ex=1$ displays late-time accelerated expansion, in complete
analogy to the DGP model.
Various phenomenological \cite{ig1,gaussbonnet} and observational \cite{igex}
consequences of the induced gravity terms have been considered. 

A common feature of all brane cosmologies is the presence in the Friedmann
equation of a contribution characterized as dark or Weyl or mirage radiation
\cite{binetruy,mirage,hebecker,brax}. In the absence of
induced gravity on a Friedmann-Robertson-Walker (FRW) brane and 
for an AdS-Schwarzschild bulk, the size of this contribution is related to the mass of the black hole \cite{kraus}.
In the general case of an arbitrary bulk content, it is the 
total integrated mass in the bulk that determines the dark radiation
\cite{Apostolopoulos:2004ic}. The term in the Friedmann equation for the
brane cosmological evolution does not, in general, scale $\sim \ell^{-4}$, where $\ell$
is the scale factor. For this reason, this term has been characterized as
\emph{generalized dark radiation} \cite{Apostolopoulos:2004ic,review}. Several
examples of non-standard scaling are known \cite{examples,radiat,bulk3}.
Also, the energy exchange between the brane and the bulk may result in a 
complicated cosmological expansion \cite{tet1,accel1,accel2}. 

Of particular interest is the possibility of late time cosmological
acceleration, with the parameter $q$ varying continuously from 
values below 1 to values above it. This implies that the effective 
parameter $w$ of the equation of state 
crosses the line $w=-1$ during the cosmological evolution.
There are indications that the observational data may favor such a scenario 
\cite{crossingw-1}, and various models 
have been proposed in order to realize it 
\cite{models,scten,model2,stringcross,copeland}. 

The purpose of the present work is to suggest an alternative simple mechanism  
within the context of the brane induced gravity. 
Assuming a FRW brane with induced gravity in the 
presence of a general bulk matter configuration, we determine the modified 
Friedmann and Raychaudhuri equations. 
This is achieved by implementing the Israel junction conditions \cite{israel} within the covariant formalism \cite{Maartens1, covariant, Maeda-Mizuno-Torii}. 
As we shall show, the generalized dark radiation plays an important role in the cosmological evolution. 
In particular, in the branch with $\ex=1$ the
evolution at late times is dominated by an effective cosmological constant. 
Additional corrections arise from
the brane matter and generalized dark radiation.
The effective equation of state of the cosmological fluid takes a form 
such that the parameter $w$ crosses the phantom 
divide $w=-1$ \emph{without} violation of the weak energy condition. 

In the next two sections we employ the covariant formalism in order to
derive the Friedmann and Raychaudhuri equations for a brane with induced gravity in the presence of
matter in the bulk. The role of the generalized dark radiation in the
expansion is discussed in section IV. Our conclusions are given in
section V.
Throughout this work the following index conventions
are used: bulk 5-dimensional indices are denoted by capital latin
letters $A,B,...=0,1,2,...,4$, and greek letters denote spacetime indices 
$\alpha ,\beta ,...=0,1,2,3$.

\section{The effect of bulk matter on a brane with induced gravity}

The presence of an induced 4-dimensional curvature term on the brane, as well
as matter in the bulk, leads to an effective action
\cite{DGP, Maeda-Mizuno-Torii}\ 
\begin{eqnarray}
S &=&\int d^{5}x\sqrt{-g}\left( \Lambda +M^{3}R+\mathcal{L}_{\text{\textsc{%
bulk}}}^{mat}\right) +  \notag \\
&&  \notag \\
&&+\int d^{4}x\sqrt{-g_4}\,\left( -V+\mathcal{L}_{\text{\textsc{brane}}%
}^{mat}+r_{c}M^{3}R_4\right) ,  \label{action1}
\end{eqnarray}
where $R$ is the curvature scalar of the 5-dimensional metric $g_{AB}$, 
$-\Lambda $ the bulk cosmological constant ($\Lambda >0$), $V$ the
brane tension, $g_{\alpha \beta }$
the induced 4-dimensional metric on the brane, $g_4$ its
determinant, and $R_4$ the corresponding curvature scalar.
The Einstein Equations (EE)
take the form 
\begin{equation}
G_{~B}^{A}=\frac{1}{2M^{3}}\left( T_{~B}^{A}+\Lambda \delta _{~B}^{A}\right),
\label{einstein}
\end{equation}
where the energy-momentum (EM) tensor 
$T_{~B}^{A}$ is 
\begin{equation}
T_{AB}=T_{AB}^{\text{\textsc{bulk}}}+\delta \left( s\right) \tau _{AB}.
\label{energy-momentum1}
\end{equation}
The term $T_{AB}^{\text{\textsc{bulk}}}$ is the bulk matter
contribution,
while $\tau _{AB}$ is the contribution from the brane located at
$s(x^A)=0$.

In order to simplify the problem we assume a $Z_{2}$ symmetry around the brane.
 The modified 4-dimensional EE can be derived by employing Israel's 
junction conditions and the Gauss equation for the extrinsic curvature of the surfaces $s(x^A)=0$ (normal to the unit 
spacelike vector $n^{A}$) \cite{israel, Maeda-Mizuno-Torii} 
\begin{equation}
G_{\alpha \beta }=\frac{1}{4M^{6}}\mathcal{S}_{\alpha \beta }-\mathcal{E}%
_{\alpha \beta }+\frac{1}{3M^{3}}\mathcal{F}_{\alpha \beta }+\frac{\Lambda }{%
4M^{3}}g_{\alpha \beta }.  
\label{branefieldequations} \end{equation}
The local quadratic corrections to the brane dynamics are represented by the
tensor 
\begin{equation}
\mathcal{S}_{\alpha \beta }=\frac{1}{12}\tau \tau _{\alpha \beta }-\frac{1}{4%
}\tau _{\alpha \gamma }\tau _{~\beta }^{\gamma }+\frac{3\tau _{\gamma \delta
}\tau ^{\gamma \delta }-\tau ^{2}}{24}g_{\alpha \beta }  \label{def1}
\end{equation}
where $\tau _{\alpha \beta }$ is defined according to 
\begin{equation}
\tau _{\alpha \beta }=T_{\alpha \beta }^{\text{\textsc{brane}}}
-Vg_{\alpha \beta }-2r_{c}M^{3}\hspace{0.15cm}G_{\alpha \beta },
\label{modifiedenergymomentum1}
\end{equation}
with $T_{\alpha \beta }^{\text{\textsc{brane}}}$ the brane matter contribution.
The presence of the induced 4-dimensional curvature
term results in a contribution 
to the tensor $\tau _{\alpha \beta }$ proportional to 
the Einstein tensor on the brane. 

The 5-dimensional effects are encoded in the projected
tensors
\begin{equation}
\mathcal{F}_{\alpha \beta }=T_{AB}^{\text{\textsc{bulk}}}g_{\hspace{0.15cm}%
\alpha }^{A}g_{\hspace{0.15cm}\beta }^{B}+\left( T_{AB}^{\text{\textsc{bulk}}%
}n^{A}n^{B}-\frac{T^{\text{\textsc{bulk}}}}{4}\right) g_{\alpha \beta }
\label{def2}
\end{equation}
\begin{equation}
\mathcal{E}_{\alpha \beta }=\mathcal{E}_{AB}g_{\hspace{0.15cm}\alpha }^{A}g_{%
\hspace{0.15cm}\beta }^{B}=C_{ACBD}n^{C}n^{D}g_{\hspace{0.15cm}\alpha
}^{A}g_{\hspace{0.15cm}\beta }^{B}.  \label{Weyl1}
\end{equation}
The tensors $\mathcal{F}_{\alpha \beta }$ and $\mathcal{E}_{\alpha \beta }$
are associated with the contributions from the matter and the free
gravitational field of the 5-dimensional bulk. We note that, in the case of
an empty bulk, the 5-dimensional contributions on the brane arise only
through the non-local effects of the free gravitational field incorporated in 
$\mathcal{E}_{\alpha \beta }$ (5D bulk gravitons).

For latter use, it is convenient to perform an 
1+3 decomposition of the above tensors into
irreducible parts with respect to the brane fluid velocity $\tilde{u}_{\alpha
}\equiv g_{\hspace{0.15cm}\alpha }^{A}$ $\tilde{u}_{A}$. The 
decomposition of the bulk EM tensor reads \cite{Apostolopoulos:2004ic} 
\begin{equation}
T_{AB}^{\text{\textsc{bulk}}}=\bar{\rho}\tilde{u}_{A}\tilde{u}_{B}+\bar{p}%
\tilde{h}_{AB}+2\bar{q}_{(A}\tilde{u}_{B)}+\bar{\pi}_{AB},
\label{energy-decomp1}
\end{equation}
where the energy density $\bar{\rho}$, isotropic pressure $\bar{p}$, energy
flux vector $\bar{q}_{A}$, and anisotropic pressure tensor $\bar{\pi}_{AB}$
are the corresponding \emph{bulk dynamical quantities} as measured by the
brane observers $\tilde{u}^{A}$. 
Here $\tilde h_{AB}=g_{AB}+\tilde u_A \tilde u_B$ is the projection tensor
normally to the prolongated brane velocity $\tilde u_A$.
A straightforward calculation gives
\begin{eqnarray}
\mathcal{F}_{\alpha \beta } &=&\frac{3\bar{\rho}+4\left( \bar{p}-\bar{p}%
_{\parallel }\right) }{4}\tilde{u}_{\alpha }\tilde{u}_{\beta }+\frac{3\bar{%
\rho}+4\left( \bar{p}-2\bar{p}_{\parallel }\right) }{12}\tilde{h}_{\alpha
\beta }+  \notag \\
&&  \notag \\
&&+2\bar{q}_{(\alpha }\tilde{u}_{\beta )}+\hat{\pi}_{\alpha \beta },
\label{BulkMatterContribution1}
\end{eqnarray}
where $\bar{p}_{\parallel }=T_{AB}^{\text{\textsc{bulk}}}n^{A}n^{B}$ is the
bulk pressure in the direction perpendicular to the brane as measured by a
brane observer \cite{Apostolopoulos:2004ic}.
The traceless tensor $\hat{\pi}_{\alpha \beta }$ is
the contribution from the 
bulk anisotropic pressure to the effective
4-dimensional gravitational field.

The Weyl ``electric'' part tensor $\mathcal{E}_{\alpha \beta }$
is decomposed as \cite{Maartens1} 
\begin{equation}
\mathcal{E}_{\alpha \beta }=\mathcal{E}\left( \tilde{u}_{\alpha }\tilde{u}%
_{\beta }+\frac{1}{3}\tilde{h}_{\alpha \beta }\right) +2\mathcal{Q}_{(\alpha
}\tilde{u}_{\beta )}+\mathcal{P}_{\alpha \beta },  
\label{WeylDecomposition}
\end{equation}
where $\mathcal{E}$, $\mathcal{Q}_{\alpha }$ and $\mathcal{P}_{\alpha \beta
} $ correspond to dynamical quantities of the
5-dimensional bulk spacetime.

With these identifications the divergence of the EM tensor of the brane matter, derived
also from the Codacci equation for the extrinsic curvature, implies 
\cite{Maartens1,Maeda-Mizuno-Torii} 
\begin{equation}
\tau _{\alpha \beta}^{\hspace{0.4cm};\beta }
=T^{\text{\textsc{brane}}\hspace{0.2cm};\beta }_{\alpha \beta }
=-2\left[ \left( \bar{q}_{C}n^{C}\right) \tilde{u}_{\alpha }+%
\bar{\pi}_{AB}n^{B}g_{\hspace{0.15cm}\alpha }^{A}\right] .
\label{Bianchi-brane}
\end{equation}
This shows that, in general, the brane matter is not conserved. There can be
energy exchange (outflow or inflow) between the brane and the bulk, 
depending 
on the character of the vector field $\left( \bar{q}_{C}n^{C}\right) 
\tilde{u}_{\alpha }+\bar{\pi}_{AB}n^{B}g_{\hspace{0.15cm}\alpha }^{A}$ 
that includes the
energy flux vector $\bar{q}_{A}$ and the bulk anisotropic stress vector 
$\bar{\pi}_{AB}n^{B}$. 

It is important to point out that the exact form or, at least, 
the set of equations that governs the evolution of the above bulk dynamical quantities as well as the rate of the energy exchange, 
cannot be deduced from \emph{on brane} considerations and one has to incorporate the full EE (\ref{einstein}).

\section{Generalized Friedmann and Raychaudhuri equations on a FRW brane}

In the presence of bulk matter, the basic tool for the covariant description
of the
brane dynamics is the irreducible decomposition (\ref
{BulkMatterContribution1}) and (\ref{WeylDecomposition}). These equations
represent the most general form of the tensors $\mathcal{F}_{\alpha \beta }$
and $\mathcal{E}_{\alpha \beta }$ and hold for any geometric brane
background. We shall restrict our considerations to a FRW brane, for which the
3-dimensional hypersurfaces $\mathcal{D}$ normal to the \emph{prolongated}
cosmological observers $\tilde{u}_{A}$ are invariant under a six-dimensional
group of isometries. It follows that the surfaces 
$\mathcal{D}$ have constant curvature,
parametrized by the constant $k_c=0,\pm 1$. 

The assumption of maximal symmetry of the spatial
hypersurfaces $\mathcal{D}$ implies that \cite{Maartens1}
\begin{equation}
\mathcal{E}_{\alpha \beta }=\mathcal{E}\left( \tilde{u}_{\alpha }\tilde{u}%
_{\beta }+\frac{1}{3}\tilde{h}_{\alpha \beta }\right) .
\label{WeylDecomposition2}
\end{equation}
In addition we have $\bar{q}_{\alpha }=0=\hat{\pi}_{\alpha \beta }$ and 
$4\bar{p}=\bar{p}_{\parallel }+3p_{\perp }$, where $p_{\perp }$ is the
isotropic pressure contribution of the bulk fluid. 
Consequently, equation (\ref{BulkMatterContribution1}) becomes 
\begin{equation}
\mathcal{F}_{\alpha \beta }=\bar{\rho}\tilde{u}_{\alpha }\tilde{u}_{\beta
}+p_{\perp }\tilde{h}_{\alpha \beta }.  \label{BulkMatterContribution2}
\end{equation}
We observe that both the bulk corrections (\ref{WeylDecomposition2}) and (%
\ref{BulkMatterContribution2}) have a perfect fluid form. In particular the
``dark fluid'' component $\mathcal{E}_{\alpha \beta }$ has a radiation
equation of state that justifies, in the case of an AdS-Schwarzschild bulk, the use of the term dark or 
Weyl radiation for its contribution to the effective Friedmann equation.

We should emphasized that the isotropy of the brane implies the existence of a
preferred spacelike direction $e^{A}$ representing the (local) axis of
symmetry with respect to which all the geometrical, kinematical and
dynamical quantities are invariant. Although the preferred spatial direction
can be chosen in two different ways, it is natural to select $e^{A}$ to be
the normal to the timelike congruence generated by the bulk observers $u^{A}$%
, i.e. $u^{A}e_{A}=0$. In this case, the radius $\ell $ represents the
average length scale for distances between any pair of brane observers and
essentially corresponds to \emph{the scale factor} of the FRW brane. It is 
defined according to \cite{Apostolopoulos:2004ic}
\begin{equation}
3\left( \ln \ell \right) _{;A}e^{A}\equiv \left( g^{AB}+u^{A}u^{B}\right)
e_{A;B}.  \label{ScaleFactorDefinition}
\end{equation}
As we have noted in section II, the 
dark radiation component $\mathcal{E}$
cannot be determined from on brane considerations because it depends on
the bulk
degrees of freedom. It follows that we must solve the 5-dimensional EE (\ref
{einstein}) in order to find the exact form of $\mathcal{E}$. This has been
done in \cite{Apostolopoulos:2004ic} for an arbitrary bulk matter configuration. In
particular we can express the dark radiation $\mathcal{E}$ in
terms of the \emph{integrated mass} of the bulk fluid within radius $\ell $:
\begin{equation}
\mathcal{E}=-\frac{1}{2M^{3}}\left[ \frac{1}{2}\left( T^{\text{\textsc{bulk%
}}}-4p_{\perp }\right) +\frac{\mathcal{M}}{\pi ^{2}\ell ^{4}}\right] ,
\label{electric1}
\end{equation}
where 
\begin{equation}
\mathcal{M}=\int_{\ell _{0}}^{\ell }2\pi ^{2}\rho \ell ^{3}d\ell +\mathcal{M}%
_{0}  \label{mass-function}
\end{equation}
is the generalized comoving mass of the bulk fluid within a spherical shell
with radii $\ell _{0}$, $\ell $, and 
$\rho =T_{AB}^{\text{\textsc{bulk}}}u^{A}u^{B}$ 
is the energy density of the bulk fluid as measured by the bulk
observers $u^{A}$. This interpretation is strictly correct only for $k_c=1$.
However, we shall refer to $\mathcal{M}$ as the integrated mass for all
geometries of the hypersurfaces $\mathcal{D}$. The integration constant 
$\mathcal{M}_{0}$ in equation (\ref{mass-function}) 
can be interpreted as the mass
of a black hole at $\ell _{0}=0$. 

Assuming a perfect fluid matter configuration on the brane, the
energy-momentum tensor can be 
written in terms of the energy density $\tilde{\rho}$ 
and the isotropic pressure $\tilde{p}$ as 
\begin{equation}
T_{\alpha \beta }^{\text{\textsc{brane}}}
=\tilde{\rho}\tilde{u}_{\alpha }\tilde{u}_{\beta }+\tilde{p}%
\tilde{h}_{\alpha \beta }.
\label{BraneEnergyMomentum}
\end{equation}
The divergence of the brane EM tensor (\ref{Bianchi-brane}) can be split along and
normally to $\tilde{u}^{\alpha }$. In this way we obtain the equation that describes the transfer of energy between the bulk and the brane 
\begin{equation}
\overset{\cdot }{\tilde{\rho}}+3H\left( \tilde{\rho}+\tilde{p}\right) =2\bar{%
q}_{C}n^{C}.
\label{exchange}
\end{equation}
The brane evolution can be studied by determining the generalized Friedmann
and Raychaudhuri equations \emph{on the brane} in the presence of bulk
matter. These follow from the Gauss-Codazzi equations and the timelike part
of the trace of the Ricci identities applied to the (irrotational, geodesic
and shear-free) timelike congruence $\tilde{u}_{\alpha }$. They have the
form 
\begin{equation}
H^{2}=\frac{1}{3}R_{\alpha \beta }\tilde{u}^{\alpha }\tilde{u}^{\beta }-%
\frac{R_3}{6}+\frac{R_4}{6}  \label{Friedmann1}
\end{equation}
\begin{equation}
\dot{H}=-H^{2}-\frac{1}{3}R_{\alpha \beta }\tilde{u}^{\alpha }\tilde{u}%
^{\beta },  \label{Raychaudhuri1}
\end{equation}
where $R_{\alpha \beta }=G_{\alpha \beta }-\frac{G}{2}g_{\alpha \beta }$ is
the modified Ricci tensor of the brane, $3H=\tilde{u}_{;\alpha }^{\alpha }$
the Hubble parameter, $R_3$ the scalar curvature of the 3-dimensional
hypersurfaces $\mathcal{D}$, and a dot denotes differentiation with respect
to $\tilde{u}^{\alpha }$, i.e. $\dot{H}\equiv H_{;\alpha }\tilde{u}^{\alpha }$.

Using equations (\ref{branefieldequations})--(\ref{modifiedenergymomentum1})
and (\ref{WeylDecomposition2})--(\ref{BraneEnergyMomentum}) we obtain
\begin{eqnarray}
H^{2} &\equiv &\left( \frac{\dot{\ell}}{\ell }\right) ^{2}=\frac{G_{\alpha
\beta }\tilde{u}^{\alpha }\tilde{u}^{\beta }}{3}-\frac{R_3}{6}=
-\frac{R_3}{6}  \notag \\
&&  \notag \\
&&+\frac{\left( \tilde{\rho}+V-2r_{c}M^{3}G_{\alpha \beta }\tilde{u}^{\alpha
}\tilde{u}^{\beta }\right) ^{2}}{144M^{6}}  \notag \\
&&  \notag \\
&&+\frac{\mathcal{M}}{6\pi ^{2}M^{3}\ell ^{4}}-\frac{\Lambda }{12M^{3}},
\label{Friedmann3}
\end{eqnarray}
\begin{eqnarray}
&&\dot{H}=-H^{2}-\frac{\Lambda }{12M^{3}}-\frac{2\left( \tilde{\rho}%
+V-2r_{c}M^{3}G_{\alpha \beta }\tilde{u}^{\alpha }\tilde{u}^{\beta }\right)
^{2}}{144M^{6}}-  \notag \\
&&  \notag \\
&&-\frac{3\left( \tilde{p}-V-2r_{c}M^{3}G_{\alpha \beta }x^{\alpha }x^{\beta
}\right) \left( \tilde{\rho}+V-2r_{c}M^{3}G_{\alpha \beta }\tilde{u}^{\alpha
}\tilde{u}^{\beta }\right) }{144M^{6}}  \notag \\
&&  \notag \\
&&-\frac{\mathcal{M}}{6\pi ^{2}M^{3}\ell ^{4}}-\frac{\bar{p}_{\parallel }}{%
6M^{3}},  \label{Raychaudhuri3}
\end{eqnarray}
where $x^{\alpha }$ is a unit spacelike vector field 
that lies on the spatial hypersurfaces ${\cal D}$ and is normal to the brane
fluid velocity vector ($x^{\alpha }x_{\alpha }=1$, 
$x^{\alpha }\tilde{u}_{\alpha }=0$).

Equation (\ref{Friedmann3}) is quadratic with respect to the timelike eigenvalue of
the induced Einstein tensor on the brane, and can be solved to obtain
\begin{eqnarray}
&&G_{\alpha \beta }\tilde{u}^{\alpha }\tilde{u}^{\beta }=\frac{V+\tilde{\rho}%
}{2M^{3}r_{c}}+\frac{6}{r_{c}^{2}}  \notag \\
&&  \notag \\
&+&\epsilon \sqrt{3}\left[\frac{12}{r_{c}^{4}}\frac{2(V+\tilde{\rho})}{%
M^{3}r_{c}^{3}}-\frac{2\mathcal{M}}{\pi ^{2}M^{3}\ell ^{4}r_{c}^{2}}+\frac{%
\Lambda }{M^{3}r_{c}^{2}}\right]^{1/2}  
\label{Einsteinzerocomponents} \end{eqnarray}
with $\epsilon=\pm 1$.
The generalized Friedmann equation becomes
\begin{eqnarray}
&&\frac{r_{c}^{2}}{2}\left( H^{2}+\frac{R_3}{6}\right) =1
+\frac{r_{c}\left( V+\tilde{\rho}\right) }{12M^{3}}  \notag \\
&&  \notag \\
&&+\epsilon \left [ 1+\frac{r_c\left( V+\tilde{\rho}\right) }{6M^{3}}+
\frac{ r_{c}^{2}\Lambda}{12M^{3}}-\frac{r_c^2 \mathcal{M}}{6\pi
^{2}M^{3}\ell ^{4}}\right]^{1/2}.  
\label{GeneralizedFriedmann3}
\end{eqnarray}
From equations (\ref{branefieldequations})--(\ref
{modifiedenergymomentum1}) and (\ref{WeylDecomposition2})--(\ref
{mass-function}) we find 
\begin{eqnarray}
G_{\alpha \beta }x^{\alpha }x^{\beta } &=&\frac{\tau _{\alpha \beta }\tilde{u%
}^{\alpha }\tilde{u}^{\beta }\left( \tau _{\alpha \beta }\tilde{u}^{\alpha }%
\tilde{u}^{\beta }+2\tau _{\alpha \beta }x^{\alpha }x^{\beta }\right) }{%
48M^{6}}+  \notag \\
&&  \notag \\
&&+\frac{\Lambda }{4M^{3}}+\frac{\mathcal{M}}{6\pi ^{2}M^{3}\ell ^{4}}
+\frac{\bar{p}_{\parallel }}{3M^{3}}.  
\label{SpatialComponent1} \end{eqnarray}
This implies
\begin{eqnarray*}
&&G_{\alpha \beta }x^{\alpha }x^{\beta }=\frac{1}{2r_{c}^{2}M^{3}G_{%
\alpha \beta }\tilde{u}^{\alpha }\tilde{u}^{\beta }+12M^{3}-r_{c}\left( 
\tilde{\rho}+V\right) }\times \\
&& \\
&&\biggl[
\frac{2\mathcal{M}}{\pi ^{2}\ell ^{4}}+\frac{\tilde{\rho}^{2}-2\tilde{p}%
\tilde{\rho}}{4M^{3}}+4\bar{p}_{\parallel }+3\Lambda \\
&& \\
&&+\left( G_{\alpha \beta }\tilde{u}^{\alpha }\tilde{u}^{\beta }\right)
^{2}M^{3}r_{c}^{2}-\left( G_{\alpha \beta }\tilde{u}^{\alpha }\tilde{u}%
^{\beta }\right) \left( 2V-\tilde{p}+\tilde{\rho}\right) r_{c} \\
&& \\
&&+\frac{3V^{2}-2V\left( \tilde{p}-2\tilde{\rho}\right) }{4M^{3}}\biggr].
\label{Ray4} \end{eqnarray*}
The r.h.s. of eq. (\ref{exchange}) accounts for the possible energy exchange
between the bulk and the brane. Because of the $Z_2$ symmetry that we have
assumed around the brane, the energy flows in or out of both sides of the
brane with equal rates. Several explicit cases of such behavior have been
studied in the past. 
In ref. \cite{bulk3} an example of a non-static bulk populated by
non-relativistic matter is given. The bulk
matter is pressureless, but has an initial outgoing velocity in the 
radial direction. The bulk metric is assumed to have the
AdS-Tolman-Bondi form. Bulk matter can flow into the brane and modify the
cosmological expansion. An interesting possibility is to identify
this type of matter with dark matter, whose density would then scale 
differently from $\ell^{-3}$.
Another interesting case is discussed in refs. \cite{radiat}.
The bulk metric is assumed to have the AdS-Vaidya form,
and the bulk energy-momentum tensor corresponds to
a radiation field. The resulting cosmological solution describes a
brane Universe that exchanges (emits or absorbs) relativistic matter 
with the bulk. In a particular application, the energy loss from the brane  
has been matched to the rate of production
of Kaluza-Klein gravitons during the collisions 
in a thermal bath of brane particles \cite{hebecker,radiat}. 

In all the above cases the distribution of matter in the bulk and the
brane is assumed to be consistent with the six-dimensional group of
isometries that permits the embedding of a 
Friedmann-Robertson-Walker brane. As a result, the possible presence of 
inhomogeneities in the distribution of brane matter is neglected.
We also note that the full description of the brane dynamics (e.g. the form
of the effective equation of state of the ``mirage" component or
the rate in which bulk matter is transformed into ``mirage" matter)
requires explicit input about
the form of the bulk matter and its interaction with the brane matter
(i.e. the use of the full 5D field equations). Nevertheless, we shall see in the next section
that the cosmological evolution can be efficiently described through the
generalized dark radiation term (generalized comoving mass) without
explicit reference to the bulk.

\section{Accelerated expansion and $w=-1$ crossing}

The Friedmann equation (\ref{GeneralizedFriedmann3}) is a 
generalization of the well known equation for an AdS-Schwarzschild bulk
\cite{deffayet}. In the standard case there is a black hole located
at $\ell_0=0$, so that the  integrated mass is constant: 
$\mathcal{M}=\mathcal{M}_0$. For this value 
our result (\ref{GeneralizedFriedmann3})
contains the known contribution characterized as dark radiation. 
For a general bulk content, the dark radiation term is replaced by the 
contribution $\mathcal{M}/(6\pi^{2}M^{3}\ell ^{4})$. 
A non-trivial bulk matter configuration leads to an 
integrated mass $\mathcal{M}$ that 
is a function of the scale factor $\ell$. As a result, the
generalized dark radiation term does not
scale $\sim \ell^{-4}$. For this reason, this term has been characterized as
\emph{generalized dark radiation} \cite{Apostolopoulos:2004ic,review}.
In references \cite{examples,radiat,bulk3} several examples were given, with
the radiation term scaling $\sim \ell^{-n}$ with $n=0,2,3$, or having more
complicated behavior. 

The two values of $\epsilon$ in equation (\ref{GeneralizedFriedmann3})
correspond to two disconnected branches of solutions. 
The nature of the predicted expansion is clearer in the limit of low
energy density. 
Let us consider first the Randall-Sundrum case \cite{rs}, in which  
the bulk cosmological constant $-\Lx$ and the 
brane tension $V$ are related through $\Lx=V^2/(12M^3)$.
We define the energy scale $k=V/(12M^3)=\left[\Lx/(12M^3)\right]^{1/2}$. 
The Friedmann equation (\ref{GeneralizedFriedmann3}) can be written as
\begin{eqnarray}
&&\frac{r_c^2}{2}\left( H^2 + \frac{k_c}{\ell^2} \right)
=1+ kr_c + kr_c \frac{\rht}{V} 
\nonumber \\
&&+ \ex \left[ 
(1+kr_c)^2+ 2kr_c \frac{\rht}{V} 
-2 (kr_c)^2\frac{\rht_d}{V}
\right]^{1/2},
\label{friedm2} \end{eqnarray}
where $\rht_d=\calm(\ell)/\left(k \pi^2 \ell^4 \right)$
is the effective energy density of the generalized dark radiation
and the constant $k_c=0,\pm 1$ parametrizes the spatial curvature of
the brane.

For $\rht,\rht_d \ll V$, keeping only the terms linear in
$\rht$, $\rht_d$, we find
\begin{eqnarray}
H^2=
&&\frac{2(1+\ex)(1+kr_c)}{r_c^2}
+\frac{1+\ex+kr_c}{kr_c(1+kr_c)}\frac{\rht}{6\left(M^3/k\right)}
\nonumber \\
&&-\frac{\ex}{1+kr_c}\frac{\rht_d}{6\left(M^3/k\right)}
-\frac{k_c}{\ell^2}.
\label{friedm3} \end{eqnarray}
For $\ex=-1$ we have 
\begin{equation}
H^2=\frac{1}{6\mpl^2}\left(\rht+\rht_d \right)-\frac{k_c}{\ell^2},
\label{friedm4} \end{equation}
with $\mpl^2=M^3(r_c+1/k)$.
The expansion is conventional, apart from the presence of the energy
density of the generalized dark radiation.

For $\ex=1$ and $kr_c \ll 1$ we obtain
\begin{equation}
H^2=\frac{4}{r_c^2}+
\frac{1}{6\mpl^2}\left(\rht-\frac{kr_c}{2}\rht_d \right)-\frac{k_c}{\ell^2},
\label{friedm5} \end{equation}
with $\mpl^2=M^3r_c/2$.
For $\ex=1$ and $kr_c \gg 1$ we have
\begin{equation}
H^2=\frac{4k}{r_c}+
\frac{1}{6\mpl^2}\left(\rht-\rht_d \right)-\frac{k_c}{\ell^2},
\label{friedm6} \end{equation}
with $\mpl^2=M^3r_c$.
In both cases an effective cosmological constant appears, despite the 
fine tuning of the bulk cosmological constant and the brane tension.
The other striking feature is the negative sign of the contribution 
proportional to the energy density of the dark radiation.

The above features also appear in the DGP model
\cite{DGP}, characterized by $\Lx=V=0$. The Friedmann
equation (\ref{GeneralizedFriedmann3}) now reads
\begin{eqnarray}
\frac{r_c^2}{2}\left( H^2 + \frac{k_c}{\ell^2} \right)
=1&+& r_c \frac{\rht}{12 M^3}  
\nonumber \\
&+& \ex \left[ 
1+\frac{r_c}{6M^3} \left(\rht-\rht_d  \right)
\right]^{1/2},
\label{friedm7} \end{eqnarray}
with
$\rht_d=r_c\calm(\ell)/\left( \pi^2 \ell^4 \right)$.
For $\ex=1$ and $\rht\ll M^3/r_c$ we have
\begin{equation}
H^2=\frac{4}{r_c^2}+
\frac{1}{6\mpl^2}\left(\rht-\frac{1}{2}\rht_d \right)-\frac{k_c}{\ell^2},
\label{friedm8} \end{equation}
with $\mpl^2=M^3r_c/2$.

It is obvious from the above that the brane cosmological expansion 
in the branch with $\ex=1$ has novel properties arising
from:
a) an effective cosmological constant, and
b) an effective negative energy density associated with the generalized
dark radiation. The second feature is \emph{not} a consequence of a violation of
the weak energy condition, as the energy density is assumed \emph{positive} both in
the bulk and on the brane. 

In the case of an AdS-Schwarzschild bulk we have $\rht_d \sim \ell^{-4}$.
At late times the contribution from the dark radiation is 
subleading to the contribution from the brane matter 
$\rht \sim \ell^{-3}$.
On the other hand, if there is a non-trivial matter configuration in
the  bulk so that $\rht_d \sim \ell^{-n}$ with $n<3$, 
the cosmological constant and the effective negative
energy density can be the leading effects.

An example with $\rht_d \sim \ell^{-2}$ is given in reference \cite{bulk3}. 
The bulk is assumed to contain a scalar field in a global monopole 
(hedgehog) configuration. The field also interacts with the
brane through a localized quadratic potential, so that the brane can
be embedded in the bulk spacetime. There is no significant energy
exchange between the brane and the bulk at low energy densities.
For large $\ell$, the dominant contribution to the 
integrated mass arises from the field kinetic term, so that 
$\calm\sim \ell^2$ and $\rht_d \sim \ell^{-2}$. 
The contribution from the brane potential is $\sim \ell^{-4}$ and, therefore,
negligible for large $\ell$. 

The evolution in the presence of a cosmological constant $\lef$ 
and a matter contribution $\rht=\rht_0 (\ell/\ell_0)^{-n}$ is determined by
\begin{equation}
\frac{1}{\elt}\frac{d\elt}{d\tt}=\left(1+\delta \frac{c^2}{\elt^n} \right)^{1/2},
\label{friedr} \end{equation}
where $\elt=\ell/\ell_0$, $\tt=(\lef/6M^2)^{1/2}t$, $c^2=\rht_0/\lef$ and
$\delta=\pm 1$.
For $\delta=1$ the solution is
\begin{equation}
\elt^{n/2}=c\sinh\left(\frac{n}{2}t+d_s\right),
\label{sol1} \end{equation}
with $d_s=\sinh^{-1}(1/c)$, while for $\delta=-1$ it is 
\begin{equation}
\elt^{n/2}=c\cosh\left(\frac{n}{2}t+d_c\right),
\label{sol2} \end{equation}
with $d_c=\cosh^{-1}(1/c)$.
The acceleration parameter $q=\ddot{\elt}\,\elt/(\dot{\elt})^2$
for $\delta=1$ is
\begin{equation}
q=1-\frac{n}{2}\left[\cosh\left(\frac{n}{2}t+d_s\right) \right]^{-2},
\label{accel1} \end{equation}
while for $\delta=-1$ it is
\begin{equation}
q=1+\frac{n}{2}\left[\sinh\left(\frac{n}{2}t+d_c\right) \right]^{-2}.
\label{accel2} \end{equation}
The Friedmann equation for the branch with $\ex=1$ can be written
in the form (\ref{friedr}) when 
one of the energy densities $\rht$, $\rht_d$ is 
negligible.
If the dominant matter contribution comes from the non-relativistic 
matter on the brane, we have $\delta=1$, $n=3$ and $q<1$.
If the dominant matter contribution comes from the generalized dark radiation
with $n<3$, we have $\delta=-1$ and $q>1$.

For a cosmological fluid 
with an equation of state $p_{\rm eff}=w\rho_{\rm eff}$ and an effective state parameter $w$,
the acceleration parameter is $q=-(1+3w)/2$. As a result, $q>1$ implies
$w<-1$, while $q<1$ implies $w>-1$. It is apparent from our discussion
above that the branch of equation 
(\ref{GeneralizedFriedmann3}) 
with $\ex=1$ can lead to an evolution during which the line
$w=-1$ is crossed. This happens if the cosmological constant is the leading
contribution, and 
the brane dark matter dominates over the generalized dark radiation 
at early times, while the reverse takes place at later times.

\section{Summary and Conclusions}

The main results of this work are the Friedmann equation 
(\ref{Friedmann3}) or (\ref{GeneralizedFriedmann3}), and
the Raychaudhuri equation (\ref{Raychaudhuri3}).
The role of the bulk matter in the Friedmann equation takes a
very simple form. The total integrated mass in the bulk up to the location
of the brane, divided by $\ell^4$, determines the
energy density of the generalized dark radiation up to a numerical factor. 
The quantity $\ell$ is
the scale factor on the brane in the Gauss normal frame, which can be
interpreted as 
the location of the brane in the bulk frame \cite{kraus,Apostolopoulos:2004ic}.
Equation (\ref{Friedmann3}) is a generalization of the Friedmann equation 
for an AdS-Schwarzschild bulk \cite{deffayet}, and is consistent with the
respective equation in the absence of the induced gravity term
\cite{Apostolopoulos:2004ic}. 
The Raychaudhuri equation is significantly more complicated than the 
respective one in the absence of induced gravity. However, it 
demonstrates that the influence of the bulk is encoded in the bulk integrated 
mass and the pressure perpendicularly to the brane as measured 
by a brane observer.

The generalized dark radiation can play an important role in the 
cosmological evolution on the brane if it scales more slowly than 
$\sim \ell^{-3}$ and there is no significant energy exchange between 
the brane and bulk.
In such a case, at late times the dark radiation would become more important 
than the brane matter,  
that is assumed to be non-relativistic and scale $\sim \ell^{-3}$.
In section IV an example was discussed in which
the generalized dark radiation scales $\sim \ell^{-2}$ \cite{bulk3} and
the energy exchange between the brane and the bulk is negligible.

The cosmological 
evolution in the $\ex=1$ branch of equation (\ref{GeneralizedFriedmann3})
is dominated by an effective cosmological constant at late times.
The brane matter and the generalized dark radiation
modify the leading exponential expansion. 
A remarkable feature is 
that the contribution from the dark radiation in the Friedmann equation
appears with a negative
sign in the $\ex=1$ branch. As a result, the presence of the bulk 
corresponds to an effective negative energy density on the brane. 
This feature is \emph{not} a consequence of a violation of
the weak energy condition, as the energy density is assumed \emph{positive} both in
the bulk and on the brane. 

An effective negative energy density accelerates the expansion on the brane
beyond the rate induced by the cosmological constant. Therefore, the 
acceleration parameter $q$ becomes larger than $1$ when the
generalized dark radiation dominates over the non-relativistic brane
matter. This was shown explicitly in section IV. 
If the brane matter dominates, the acceleration parameter is smaller than 
$1$. It is then obvious that the value $q=1$ is obtained 
at the point of comparable size of the two contributions.
This corresponds to the crossing of the phantom divide $w=-1$ for the
effective equation of state of the cosmological fluid.

The $w=-1$ crossing is difficult to realize in dark energy models without
the inclusion of higher derivative terms or multiple fields \cite{copeland}.
In this respect, a brane Universe with induced gravity provides in a simple
way two 
desirable features: a) an effective cosmological constant, and b) an
effective negative energy density, that can trigger the $w=-1$ crossing when
it becomes larger than the energy density of the ordinary non-relativistic 
matter. 

Before concluding we must point out that a number of recent works has shown
that the self-accelerating branch of the induced gravity models contains a 
ghost mode that renders it unstable \cite{instability,lue,charmousis}. 
This is a serious problem for the physical relevance of cosmologies within this 
branch. It may still be possible to construct a viable model if the instability is
not rapid \cite{lue}. An alternative possibility is that the ghost modes are eliminated
by an additional mechanism, such as the compactification of the extra dimension 
\cite{charmousis}. For example, this may be feasible in the context of the Randall-Sundrum model,
if the effective compactification radius $1/k$ is chosen appropriately within the range $1/k<r_c$.
The resulting cosmological evolution is described by eq. (\ref{friedm6}), in which an effective
cosmological constant and a generalized radiation term with a negative sign appear.
As remarked in \cite{charmousis}, an arbitrariness in the fine tuning of the cosmological constant is implicit
in such a model. In the DGP model \cite{DGP} the brane tension and the bulk cosmological constant are taken equal to zero, 
while in our example they
are chosen similarly to the Randall-Sundrum model, so as to give contributions that cancel each other \cite{rs}.

\section{Acknowledgments}

This work was supported through the research program 
``Pythagoras II'' (grant 70-03-7992)
of the Greek Ministry of National Education, 
partially funded by the European Union.

\end{document}